\documentclass[english,aps,superscriptaddress,preprintnumbers]{revtex4}
\usepackage[T1]{fontenc}
\usepackage[latin9]{inputenc}
\usepackage{mathrsfs}
\usepackage{amsmath}
\usepackage{amssymb}
\usepackage{esint}

\makeatletter
\@ifundefined{textcolor}{}
{%
 \definecolor{BLACK}{gray}{0}
 \definecolor{WHITE}{gray}{1}
 \definecolor{RED}{rgb}{1,0,0}
 \definecolor{GREEN}{rgb}{0,1,0}
 \definecolor{BLUE}{rgb}{0,0,1}
 \definecolor{CYAN}{cmyk}{1,0,0,0}
 \definecolor{MAGENTA}{cmyk}{0,1,0,0}
 \definecolor{YELLOW}{cmyk}{0,0,1,0}
}

\usepackage{mathrsfs}

\@ifundefined{textcolor}{}{%
 \definecolor{BLACK}{gray}{0}
 \definecolor{WHITE}{gray}{1}
 \definecolor{RED}{rgb}{1,0,0}
 \definecolor{GREEN}{rgb}{0,1,0}
 \definecolor{BLUE}{rgb}{0,0,1}
 \definecolor{CYAN}{cmyk}{1,0,0,0}
 \definecolor{MAGENTA}{cmyk}{0,1,0,0}
 \definecolor{YELLOW}{cmyk}{0,0,1,0}
}

\usepackage{babel}

\usepackage{babel}

\usepackage{babel}

\usepackage{babel}

\makeatother

\usepackage{babel}
\begin{document}
\title{Chiral magnetic effect for chiral fermion system }
\author{Ren-Da Dong}
\affiliation{Institute of Particle Physics and Key Laboratory of Quark and Lepton
Physics (MOS), Central China Normal University, Wuhan 430079, China}
\author{Ren-Hong Fang}
\email{renhong.fang@mail.ccnu.edu.cn}

\affiliation{Institute of Particle Physics and Key Laboratory of Quark and Lepton
Physics (MOS), Central China Normal University, Wuhan 430079, China}
\author{De-Fu Hou}
\email{houdf@mail.ccnu.edu.cn}

\affiliation{Institute of Particle Physics and Key Laboratory of Quark and Lepton
Physics (MOS), Central China Normal University, Wuhan 430079, China}
\author{Duan She}
\affiliation{Institute of Particle Physics and Key Laboratory of Quark and Lepton
Physics (MOS), Central China Normal University, Wuhan 430079, China}
\begin{abstract}
We concisely derive chiral magnetic effect through Wigner function
approach for chiral fermion system. Then we derive chiral magnetic
effect through solving the Landau levels of chiral fermions in detail.
The procedures of second quantization and ensemble average lead to
the equation of chiral magnetic effect for righthand and lefthand
fermion systems. Chiral magnetic effect only comes from the contribution
of the lowest Landau level. We carefully analyze the lowest Landau
level, and find that all righthand (chirality is $+1$) fermions move
along positive $z$-direction and all lefthand (chirality is $-1$)
fermions move along negative $z$-direction. From this picture chiral
magnetic effect can be explained clearly in a microscopic way. 
\end{abstract}
\maketitle

\section{Introduction}

Quark gluon plasma (QGP) can be created in high energy heavy ion collisions,
which is a extremely hot and dense matter. Very huge magnetic field
can be produced for high energy peripheral collisions \citep{Deng:2012pc,Tuchin:2014iua,Li:2016tel}.
One of the predictions in QGP is that positively charged particles
and negatively charged particles will seperate along the direction
of magnetic field, which is related to chiral magnetic effect (CME)
\citep{Kharzeev:2004ey,Kharzeev:2007jp,Kharzeev:2007tn}. Many efforts
has been made to find the signal of CME in experiments \citep{Abelev:2009ac,Abelev:2009ad,Abelev:2012pa}.
But due to the background noise, no definite CME singal has been found.
There are also many theoretical methods to study CME, such as AdS/CFT
\citep{Erdmenger:2008rm,Kalaydzhyan:2011vx}, hydrodynamics \citep{Son:2009tf,Pu:2010as,Kharzeev:2011ds},
finite temperature field theory \citep{Fukushima:2008xe,Wu:2016dam,Metlitski:2005pr,Miransky:2015ava},
quantum kinetic theory \citep{Gao:2012ix}, et al.

In this article we will carefully study CME through solving Landau
levels. For massive Dirac fermion system, there were some works on
CME related to Landau levels. In their paper \citep{Fukushima:2008xe},
Fukushima et al. proposed four methods to derive CME, in one of which
they took use of Landau energy levels for massive Dirac equation with
chemical potential $\mu$ and chiral chemical potential $\mu_{5}$
in a homogeneous magnetic background $\boldsymbol{B}=B\boldsymbol{e}_{z}$
to construct the thermodynamic potential $\Omega$. The macroscopic
electric current $j^{z}$ along $z$-axis can be obtained from the
thermodynamic potential $\Omega$. Another work on CME through Landau
levels is related to the second quantization of Dirac field. In their
work \citep{Sheng:2017lfu}, the authors solved the Landau levels
and corresponding Landau wavefunctions for massive Dirac equation
in a background of uniform magnetic field also with chemical potential
$\mu$ and chiral chemical potential $\mu_{5}$. Then they secondly
quantized Dirac field and expanded it by these solved Landau wavefunctions
and creation/construction operators. The density operator $\hat{\rho}$
can then be determined from Hamiltonian $\hat{H}$ and particle number
operator $\hat{N}$ of the system. Finally they derived the macroscopic
electric current $j^{z}$ along $z$-axis through the trace of density
operator $\hat{\rho}$ and electric current operator $\hat{j}^{z}$,
which is just CME equation.

From the study on CME for massive Dirac fermions through Landau levels,
one can conclude that the contribution to CME only come from the lowest
Landau level, while the contributions from high Landau levels cancel
with each other. However, due to the mass $m$ of Dirac fermion, the
physical picture of CME for massive Dirac fermion system is not as
clear as massless fermion case, because the physical meaning of chiral
chemical potential $\mu_{5}$ for massive fermion case is not very
clear in fact. To see this issue clearly, we list the lowest Landau
level as follows (We set the homogeneous magnetic background $\boldsymbol{B}=B\boldsymbol{e}_{z}$
along $z$-axis and assume $eB>0$, which is also appropriate for
following sections), 
\begin{equation}
\psi_{0\lambda}(k_{y},k_{z};\boldsymbol{x})=c_{0\lambda}\left(\begin{array}{c}
\varphi_{0}\\
0\\
F_{0\lambda}\varphi_{0}\\
0
\end{array}\right)\frac{1}{L}e^{i(yk_{y}+zk_{z})},\ (\lambda=\pm1),\label{eq:intro-3}
\end{equation}
with energy $E=\lambda\sqrt{m^{2}+k_{z}^{2}}$, where $F_{0\lambda}=(\lambda\sqrt{m^{2}+k_{z}^{2}}+k_{z})/m$,
and $\varphi_{0}$ is the zeroth harmonic oscillator wavefunction
along $x$-axis. To simplify following discussions, we can set $k_{y}=0$.
The $z$-component of spin operator for single particle is $S^{z}=\frac{1}{2}\text{diag}\,(\sigma_{3},\sigma_{3})$,
which implies $S^{z}\psi_{0\lambda}=(+\frac{1}{2})\psi_{0\lambda}$.
When $\lambda=+1$, $E=\sqrt{m^{2}+k_{z}^{2}}>0$, then $\psi_{0+}$
in Eq. (\ref{eq:intro-3}) describes a particle with momentum $k_{z}$
and spin projection $S^{z}=+\frac{1}{2}$. When $\lambda=-1$, $E=-\sqrt{m^{2}+k_{z}^{2}}<0$,
then $\psi_{0-}$ in Eq. (\ref{eq:intro-3}) describes an antiparticle
with momentum $-k_{z}$ and spin projection $S^{z}=-\frac{1}{2}$.
So in the homogeneous magnetic background $\boldsymbol{B}=B\boldsymbol{e}_{z}$,
we obtain a picture for the lowest Landau level (with $k_{y}=0$):
all particles spin along $(+z)$-axis while all antiparticles spin
along $(-z)$-axis, but the $z$-component momentum of particles and
anti-particles can be along $(+z)$-axis or $(-z)$-axis. In fact
it is very difficult to obtain a net electric current along the magnetic
field direction from the picture of the lowest Landau level for massive
fermion case.

In this article, we focus on massless fermion (also called ``chiral
fermion'') system, in which case we will show that it is very easy
to obtain a net electric current along the magnetic field direction
from the picture of the lowest Landau level. Chiral fermion field
can be divided into two independent parts --- righthand part and
lefthand part. Firstly let us set up notation. The electric charge
of a fermion/antifermion is $\pm e$. The chemical potential for righthand/lefthand
fermions is $\mu_{R/L}$, from which chiral chemical potential and
the ordinary chemical potential can be expressed as $\mu_{5}=(\mu_{R}-\mu_{L})/2$,
$\mu=(\mu_{R}+\mu_{L})/2$. The chemical potential $\mu$ describes
the imbalance of fermions and anti-fermions, while the chiral chemical
potential $\mu_{5}$ describes the imbalance of righthand and lefthand
chirality. It is worth noting that an introduction of a chemical potential
generally correspond to a conserved quantity. The conserved quantity
corresponding to the ordinary chemical potential $\mu$ is total electric
charge of the system. But due to chiral anomaly \citep{Adler:1969gk,Bell:1969ts},
there is no conserved quantity corresponding to the chiral chemical
potential $\mu_{5}$, which is crucial for the existence of CME \citep{Kharzeev:2013ffa}.

To study CME for chiral fermion system, firstly we show a succinct
derivation of CME through Wigner function approach, from which we
can obtain CME as a quantum effect of the first order in $\hbar$
expansion. Then we turn to solve the Landau levels for the chiral
fermion system. Since chiral fermions are massless, the equations
of righthand and lefthand parts of the chiral fermion field decouple
with each other, which allow us to deal with righthand and lefthand
fermion fields independently. Taking righthand fermion field as an
example, we firstly solve the energy eigenvalue equation of righthand
fermion field in an external uniform magnetic field, and obtain a
series of Landau levels. Then we perform the second quantization for
righthand fermion field which can be expanded by the complete wavefunctions
of Landau levels. Finally chiral magnetic effect can be derived through
ensemble average, from which we see explicitly that CME only comes
from the lowest Landau level. By analyzing the physical picture for
the lowest Landau level, we conclude that all righthand (chirality
is $+1$) fermions move along positive $z$-direction and all lefthand
(chirality is $-1$) fermions move along negative $z$-direction.
This is the main result of this article. This result can qualitatively
explain why there is a macroscopic electric current along the direction
of the magnetic field in a chiral fermion system, which is called
CME. We amphasize that CME equation is derived from solving Landau
levels without the approaximation of weak magnetic field.

The rest of this article is organized as follows. In Sec. \ref{sec:Wigner-function},
we give a succinct derivation for CME through Wigner function approach.
In Sec. \ref{sec:Hamiltonian}, we solve the Landau levels for righthand
fermion field. In Sec. \ref{sec:Second-quantization} and \ref{sec:Chiral-magnetic-effect},
we perform second quantization of the righthand fermion system and
obtain CME through ensemble average. In Sec. \ref{sec:Physical-picture},
we discuss the physical picture of the lowest Landau level. At last,
we summarize this aiticle in Sec. \ref{sec:Summary}. We present some
of derivation details in the appendixes.

Throughout this article we adopt natural units where $\hbar=c=k_{B}=1$.
The convention for the metric tensor is $g^{\mu\nu}=\text{diag\,}(+1,-1,-1,-1)$.
The totally antisymmetric Levi-Civita tensor is $\epsilon^{\mu\nu\rho\sigma}$
with $\epsilon^{0123}=+1$ which agrees with Peskin \citep{Peskin}
but not with Bjorken and Drell \citep{Bjorken}. Greek indices, such
as $\mu,\nu,\rho,\sigma$, run over $0,1,2,3$, or $t,x,y,z$, while
Roman indices, such as $i,j,k$, run over $1,2,3$ or $x,y,z$. We
use the Heaviside-Lorentz convention for electromagnetism.

\section{A succinct derivation of CME from Wigner function approach}

\label{sec:Wigner-function}In this section we will concisely derive
CME from Wigner function approach for chiral fermion system. Our starting
point is the following covariant and gauge invariant Wigner function,
\begin{equation}
\mathscr{W}_{\alpha\beta}(x,p)=\bigg\langle:\frac{1}{(2\pi)^{4}}\int d^{4}ye^{-ip\cdot y}\overline{\Psi}_{\beta}(x+\frac{y}{2})U(x+\frac{y}{2},x-\frac{y}{2})\Psi_{\alpha}(x-\frac{y}{2}):\bigg\rangle,\label{eq:oo1}
\end{equation}
where $\langle:\cdots:\rangle$ represents ensemble average, $\Psi(x)$
is the Dirac filed operator for chiral fermions, $\alpha$, $\beta$
are Dirac spinor indices, and $U(x+y/2,x-y/2)$ is the gauge link
of a straight line from $(x-y/2)$ to $(x+y/2)$. This specific choice
for the path in gauge link in the definition of Wigner function is
firstly proposed in \citep{Vasak:1987um}, where the authors argued
that this type of gauge link can make the variable $p$ in Wigner
function $\mathscr{W}(x,p)$ becoming a kinetic momentum, although
in principle the path in the gauge link is arbitrary. The specific
choice of the two points $(x\pm y/2)$ in the integrand in Eq. (\ref{eq:oo1})
is based on the consideration of symmetry. In fact we can also replace
$(x\pm y/2)$ by $(x+sy)$ and $(x-(1-s)y)$ where $s$ is a real
parameter \citep{Wigner:1932eb}.

Suppose that the electromagnetic field $F^{\mu\nu}$ is homogeneous
in space and time, then from the dynamical equation satisfied by $\Psi(x)$,
one can obtain the dynamical equation for $\mathscr{W}(x,p)$ as follows,
\begin{equation}
\gamma\cdot K\mathscr{W}(x,p)=0,\label{eq:oo2}
\end{equation}
where $K_{\mu}=\frac{i}{2}\nabla_{\mu}+p_{\mu}$ and $\nabla_{\mu}=\partial_{\mu}^{x}-eF_{\mu\nu}\partial_{p}^{\nu}$.
Since $\mathscr{W}(x,p)$ is a $4\times4$ matrix, we can decompose
it by the 16 independent $\Gamma$-matrices, 
\begin{equation}
\mathscr{W}=\frac{1}{4}(\mathscr{F}+i\gamma^{5}\mathscr{P}+\gamma^{\mu}\mathscr{V}_{\mu}+\gamma^{5}\gamma^{\mu}\mathscr{A}_{\mu}+\frac{1}{2}\sigma^{\mu\nu}\mathscr{S}_{\mu\nu}).\label{eq:oo3}
\end{equation}
The 16 coefficient functions $\mathscr{F},\mathscr{P},\mathscr{V}_{\mu},\mathscr{A}_{\mu},\mathscr{S}_{\mu\nu}$
are scalar, pseudoscalar, vector, pseudovector and tensor, respectively,
and they are all real functions due to the fact that $\mathscr{W}^{\dagger}=\gamma^{0}\mathscr{W}\gamma^{0}$.
Vector current and axial vector current can be expressed as the 4-momentum
integration of $\mathscr{V}^{\mu}$ and $\mathscr{A}^{\mu}$, 
\begin{equation}
J_{V}^{\mu}(x)=\int d^{4}p\mathscr{V}^{\mu},\label{eq:004}
\end{equation}
\begin{equation}
J_{A}^{\mu}(x)=\int d^{4}p\mathscr{A}^{\mu}.\label{eq:005}
\end{equation}

If Eq. (\ref{eq:oo2}) is multiplied by $\gamma\cdot K$ from the
lefthand side, we can obtain the quadratic form of Eq. (\ref{eq:oo2})
as follows, 
\begin{equation}
\bigg(K^{2}-\frac{i}{2}\sigma^{\mu\nu}[K_{\mu},K_{\nu}]\bigg)\mathscr{W}=0.\label{eq:oo6}
\end{equation}
From Eq. (\ref{eq:oo6}) we can obtain two off mass-shell equations
for $\mathscr{V}_{\mu}$ and $\mathscr{A}_{\mu}$ (See Appendix \ref{sec:Vlasov-equation}
for details), 
\begin{eqnarray}
(p^{2}-\frac{1}{4}\hbar^{2}\nabla^{2})\mathscr{V}_{\mu} & = & -e\hbar\tilde{F}_{\mu\nu}\mathscr{A}^{\nu},\label{eq:ap33}\\
(p^{2}-\frac{1}{4}\hbar^{2}\nabla^{2})\mathscr{A}_{\mu} & = & -e\hbar\tilde{F}_{\mu\nu}\mathscr{V}^{\nu},\label{eq:ap34}
\end{eqnarray}
where $\tilde{F}_{\mu\nu}=\frac{1}{2}\varepsilon_{\mu\nu\rho\sigma}F^{\rho\sigma}$.
Note that we have explicitly shown $\hbar$ factor in Eqs. (\ref{eq:ap33},
\ref{eq:ap34}). If we expand $\mathscr{V}^{\mu}$ and $\mathscr{A}^{\mu}$
order by order in $\hbar$ as 
\begin{align}
\mathscr{V}^{\mu} & =\mathscr{V}_{(0)}^{\mu}+\hbar\mathscr{V}_{(1)}^{\mu}+\hbar^{2}\mathscr{V}_{(2)}^{\mu}+\cdots,\label{eq:ap35}\\
\mathscr{A}^{\mu} & =\mathscr{A}_{(0)}^{\mu}+\hbar\mathscr{A}_{(1)}^{\mu}+\hbar^{2}\mathscr{A}_{(2)}^{\mu}+\cdots,\label{eq:ap36}
\end{align}
then at order $o(1)$ and $o(\hbar)$, Eqs. (\ref{eq:ap33}, \ref{eq:ap34})
become 
\begin{align}
p^{2}\mathscr{V}_{(0)}^{\mu} & =0,\label{eq:ap37}\\
p^{2}\mathscr{A}_{(0)}^{\mu} & =0,\label{eq:ap38}\\
p^{2}\mathscr{V}_{(1)\mu} & =-e\hbar\tilde{F}_{\mu\nu}\mathscr{A}_{(0)}^{\nu},\label{eq:ap39}\\
p^{2}\mathscr{A}_{(1)\mu} & =-e\hbar\tilde{F}_{\mu\nu}\mathscr{V}_{(0)}^{\nu}.\label{eq:ap40}
\end{align}

The zeroth order solutions $\mathscr{V}_{(0)}^{\mu}$ and $\mathscr{A}_{(0)}^{\mu}$
can be derived by directly calculating Wigner function without gauge
link through ensemble average in Eq. (\ref{eq:oo1}), which has already
been obtained by one of the authors and his collaborators \citep{Fang:2016vpj}.
The results for $\mathscr{V}_{(0)}^{\mu}$ and $\mathscr{A}_{(0)}^{\mu}$
are 
\begin{align}
\mathscr{V}_{(0)}^{\mu}= & \frac{2}{(2\pi)^{3}}p^{\mu}\delta(p^{2})\sum_{s}\bigg[\theta(p^{0})\frac{1}{e^{\beta(p^{0}-\mu_{s})}+1}+\theta(-p^{0})\frac{1}{e^{\beta(-p^{0}+\mu_{s})}+1}\bigg],\label{eq:ap52}
\end{align}
\begin{align}
\mathscr{A}_{(0)}^{\mu}= & \frac{2}{(2\pi)^{3}}p^{\mu}\delta(p^{2})\sum_{s}s\bigg[\theta(p^{0})\frac{1}{e^{\beta(p^{0}-\mu_{s})}+1}+\theta(-p^{0})\frac{1}{e^{\beta(-p^{0}+\mu_{s})}+1}\bigg],\label{eq:ap53}
\end{align}
where $\beta=1/T$ is the inverse temperature of the system, $\mu_{R/L}$
is the chemical potential for righthand/lefthand fermions as mentioned
in the introduction, and $s=\pm1$ corresponds to the chirality of
righthand/lefthand fermions respectively. Obviously the zeroth order
solusions $\mathscr{V}_{(0)}^{\mu}$ and $\mathscr{A}_{(0)}^{\mu}$
satisfy Eqs. (\ref{eq:ap37}, \ref{eq:ap38}), which means they are
both on shell. From Eqs. (\ref{eq:ap39}, \ref{eq:ap40}) we directly
obtain the first order solutions, 
\begin{align}
\mathscr{V}_{(1)}^{\mu}= & \frac{2}{(2\pi)^{3}}e\hbar\tilde{F}^{\mu\nu}p_{\nu}\delta^{\prime}(p^{2})\sum_{s}s\bigg[\theta(p^{0})\frac{1}{e^{\beta(p^{0}-\mu_{s})}+1}+\theta(-p^{0})\frac{1}{e^{\beta(-p^{0}+\mu_{s})}+1}\bigg],\label{eq:ap54}
\end{align}
\begin{align}
\mathscr{A}_{(1)}^{\mu}= & \frac{2}{(2\pi)^{3}}e\hbar\tilde{F}^{\mu\nu}p_{\nu}\delta^{\prime}(p^{2})\sum_{s}\bigg[\theta(p^{0})\frac{1}{e^{\beta(p^{0}-\mu_{s})}+1}+\theta(-p^{0})\frac{1}{e^{\beta(-p^{0}+\mu_{s})}+1}\bigg],\label{eq:ap55}
\end{align}
where we have used $\delta^{\prime}(p^{2})=-\delta(p^{2})/p^{2}$.
Eqs. (\ref{eq:ap54}, \ref{eq:ap55}) are the same as the second term
in Eq. (3) in \citep{Gao:2015zka}.

Now we can calculate $\boldsymbol{J}_{A/V}$ based on Eqs. (\ref{eq:004},
\ref{eq:005}). Since $\mathscr{V}_{(0)}^{i},\mathscr{A}_{(0)}^{i}$
are odd functions of 3-momentum $\boldsymbol{p}$, the nonzero contribution
to $J_{V/A}^{i}$ only comes from $\mathscr{V}_{(1)}^{i}$ and $\mathscr{A}_{(1)}^{i}$.
We assume there only exists a uniform magnetic field $\boldsymbol{B}=B\boldsymbol{e}_{z}$,
i.e. $F^{12}=-F^{21}=-B$ and $\tilde{F}^{03}=-\tilde{F}^{30}=-B$
(other components of $F^{\mu\nu}$, $\tilde{F}^{\mu\nu}$are zero),
which implies $J_{V/A}^{x}=J_{V/A}^{y}=0$. After integration over
the $z$-components of Eqs. (\ref{eq:ap54}, \ref{eq:ap55}) we have
\begin{equation}
J_{V}^{z}=\int d^{4}p\mathscr{V}_{(1)}^{z}=\frac{e\hbar\mu_{5}}{2\pi^{2}}B,\label{eq:ap45}
\end{equation}
\begin{equation}
J_{A}^{z}=\int d^{4}p\mathscr{A}_{(1)}^{z}=\frac{e\hbar\mu}{2\pi^{2}}B,\label{eq:ap46}
\end{equation}
where $\mu_{5}=(\mu_{R}-\mu_{L})/2$ and $\mu=(\mu_{R}+\mu_{L})/2$.
Eq. (\ref{eq:ap45}) means that if $\mu_{5}\neq0$, then there will
be a current along the magnetic direction. Since $\hbar$ appears
in the coefficient of the magnetic field $B$, we need a very huge
magnetic field to produce a macroscopic current, which may be realised
in high energy heavy ion collisions. So far we have derived CME for
chiral fermion system through Wigner function approach, and we can
see that CME is the first order quantum effect in $\hbar$. In fact,
Wigner function approach is a type of quantum kinetic theory, which
can imply the quantum effect of a multi-particle system, such as CME.

\section{Landau levels for righthand fermions}

\label{sec:Hamiltonian}In this section and following sections we
will derive CME for a chiral fermion system through solving Landau
levels. The Lagrangian for a chiral fermion field is 
\begin{equation}
\mathcal{L}=\overline{\Psi}(x)i\gamma\cdot D\Psi(x),\label{eq:tt3}
\end{equation}
with the covariant derivative $D^{\mu}=\partial^{\mu}+ieA^{\mu}$,
and the electric charge $\pm e$ for particles/antiparticles. For
a uniform magnetic field $\boldsymbol{B}=B\mathbf{e}_{z}$ along $z$-axis,
we can choose the gauge potential as $A^{\mu}=(0,0,Bx,0)$. The equation
of motion for the field $\Psi(x)$ is 
\begin{equation}
i\gamma\cdot D\Psi(x)=0,\label{eq:tt4}
\end{equation}
which can be written as a form of Schrödinger equation, 
\begin{equation}
i\frac{\partial}{\partial t}\Psi(t,\boldsymbol{x})=i\boldsymbol{\alpha}\cdot\boldsymbol{D}\Psi(t,\boldsymbol{x}),\label{eq:tt2}
\end{equation}
with $\boldsymbol{D}=-\nabla+ie\boldsymbol{A}$, $\boldsymbol{A}=(0,Bx,0)$.
In the chiral representation of Dirac $\gamma$-matrices where $\gamma^{5}=\text{diag}\,(-1,1)$,
$\boldsymbol{\alpha}=\text{diag}\,(-\boldsymbol{\sigma},\boldsymbol{\sigma})$,
we can write $\Psi$ in this form: $\Psi=(\Psi_{L}^{T},\Psi_{R}^{T})^{T}$.
Then Eq. (\ref{eq:tt2}) becomes 
\begin{equation}
i\frac{\partial}{\partial t}\left(\begin{array}{c}
\Psi_{L}(t,\boldsymbol{x})\\
\Psi_{R}(t,\boldsymbol{x})
\end{array}\right)=\left(\begin{array}{c}
-i\boldsymbol{\sigma}\cdot\boldsymbol{D}\Psi_{L}(t,\boldsymbol{x})\\
i\boldsymbol{\sigma}\cdot\boldsymbol{D}\Psi_{R}(t,\boldsymbol{x})
\end{array}\right),\label{eq:tt5}
\end{equation}
which indicates that the two fields $\Psi_{L/R}$, which correspond
to eigenvalues $\mp1$ of the matrix $\gamma^{5}$, decouple with
each other. The two fields $\Psi_{L/R}$ are often called lefthand/righthand
fermion fields respectively. Lefthand and righthand fermions are also
called chiral fermions.

In the following we will focus on solving the eigenvalue equation
for righthand fermion field $\Psi_{R}$ (Similar results can be obtained
for lefthand fermion field $\Psi_{L}$).

In order to obtain Landau levels, we must solve the eigenvalue equation
for righthand fermion field as follows, 
\begin{equation}
i\boldsymbol{\sigma}\cdot\boldsymbol{D}\psi_{R}=E\psi_{R},\label{eq:a}
\end{equation}
with $\boldsymbol{D}=(-\partial_{x},-\partial_{y}+ieBx,-\partial_{z})$.
The details for solving Eq. (\ref{eq:a}) are put in Appendix \ref{sec:Landau levels-1}.
We list the eigenfunctions and eigenvalues in the following : For
$n=0$ Landau level, the wavefunction with energy $E=k_{z}$ is

\begin{equation}
\psi_{R0}(k_{y},k_{z};\boldsymbol{x})=\left(\begin{array}{c}
\varphi_{0}(\xi)\\
0
\end{array}\right)\frac{1}{L}e^{i(yk_{y}+zk_{z})},\label{eq:tt6}
\end{equation}
For $n>0$ Landau level, the wavefunction with energy $E=\lambda E_{n}(k_{z})$
is

\begin{equation}
\psi_{Rn\lambda}(k_{y},k_{z};\boldsymbol{x})=c_{n\lambda}\left(\begin{array}{c}
\varphi_{n}(\xi)\\
iF_{n\lambda}\varphi_{n-1}(\xi)
\end{array}\right)\frac{1}{L}e^{i(yk_{y}+zk_{z})},\label{eq:tt7}
\end{equation}
where $\lambda=\pm1$, $E_{n}(k_{z})=\sqrt{2neB+k_{z}^{2}}$, $F_{n\lambda}(k_{z})=[k_{z}-\lambda E_{n}(k_{z})]/\sqrt{2neB}$,
normalised coefficient $|c_{n\lambda}|^{2}=1/(1+F_{n\lambda}^{2})$,
and $\varphi_{n}(\xi)=\varphi_{n}(\sqrt{eB}x-k_{y}/\sqrt{eB})$ is
the $n$-th order wavefunction of a harmonic oscillator.

For $n>0$ Landau levels, the wavefunctions with energys $E=\pm E_{n}(k_{z})$
corresponds to fermions and antifermions respecttively. For the lowest
Landau level, the wavefunction with energy $E=k_{z}>0$ corresponds
to fermions and with energy $E=k_{z}<0$ corresponds to antifermions
respecttively. Wavefunctions of all Landau levels are orthonormal
and complete. For lefthand fermion field, the eigenfunctions of Landau
levels are the same as the righthand case but with the sign of the
eigenvalues changed.

\section{Second quantization for righthand fermion field}

\label{sec:Second-quantization}In this section, we secondly quantize
the righthand fermion field $\Psi_{R}(\boldsymbol{x})$, then the
righthand fermion field $\Psi_{R}(\boldsymbol{x})$ becomes an operator
and satisfies following anticommutative relations, 
\begin{align}
\{\Psi_{R}(\boldsymbol{x}),\Psi_{R}^{\dagger}(\boldsymbol{x}^{\prime})\} & =\delta^{(3)}(\boldsymbol{x}-\boldsymbol{x}^{\prime}),\nonumber \\
\{\Psi_{R}(\boldsymbol{x}),\Psi_{R}(\boldsymbol{x}^{\prime})\} & =0.\label{eq:ss3}
\end{align}
Since all eigenfunctions for the Hamiltanian of the righthand fermion
field are orthonormal and complete, we can decompose the righthand
fermion field operator $\Psi_{R}(\boldsymbol{x})$ by these eigenfunctions
as

\begin{eqnarray}
\Psi_{R}(\boldsymbol{x}) & = & \sum_{k_{y},k_{z}}[\theta(k_{z})a_{0}(k_{y},k_{z})\psi_{R0}(k_{y},k_{z};\boldsymbol{x})+\theta(-k_{z})b_{0}^{\dagger}(k_{y},k_{z})\psi_{R0}(k_{y},k_{z};\boldsymbol{x})]\nonumber \\
 &  & +\sum_{n,k_{y},k_{z}}[a_{n}(k_{y},k_{z})\psi_{Rn+}(k_{y},k_{z};\boldsymbol{x})+b_{n}^{\dagger}(k_{y},k_{z})\psi_{Rn-}(k_{y},k_{z};\boldsymbol{x})].\label{eq:ss2}
\end{eqnarray}
Different from the general Fourier decomposition for second quantization,
we have put two theta functions $\theta(\pm k_{z})$ in front of $a_{0}(k_{y},k_{z})$
and $b_{0}^{\dagger}(k_{y},k_{z})$ in the decomposition, which is
very important for the subsequent procedure of second quantization.
From formula (\ref{eq:ss3}), we can obtain following anticommutative
relations, 
\begin{align}
\{\theta(k_{z})a_{0}(k_{y},k_{z}),\theta(k_{z}^{\prime})a_{0}^{\dagger}(k_{y}^{\prime},k_{z}^{\prime})\} & =\theta(k_{z})\delta_{k_{y}k_{y}^{\prime}}\delta_{k_{z}k_{z}^{\prime}}\nonumber \\
\{\theta(-k_{z})b_{0}(k_{y},k_{z}),\theta(-k_{z}^{\prime})b_{0}^{\dagger}(k_{y}^{\prime},k_{z}^{\prime})\} & =\theta(-k_{z})\delta_{k_{y}k_{y}^{\prime}}\delta_{k_{z}k_{z}^{\prime}}\nonumber \\
\{a_{n}(k_{y},k_{z}),a_{n^{\prime}}^{\dagger}(k_{y}^{\prime},k_{z}^{\prime})\} & =\delta_{nn^{\prime}}\delta_{k_{y}k_{y}^{\prime}}\delta_{k_{z}k_{z}^{\prime}}\nonumber \\
\{b_{n}(k_{y},k_{z}),b_{n^{\prime}}^{\dagger}(k_{y}^{\prime},k_{z}^{\prime})\} & =\delta_{nn^{\prime}}\delta_{k_{y}k_{y}^{\prime}}\delta_{k_{z}k_{z}^{\prime}}.\label{eq:tt19}
\end{align}
Note that the two theta functions $\theta(\pm k_{z})$ are always
attached to the lowest Landau level operators such as $a_{0},a_{0}^{\dagger},b_{0},b_{0}^{\dagger}$.
The Hamiltonian and total particle number of the righthand fermion
system are 
\begin{eqnarray}
H & = & \int d^{3}x\Psi_{R}^{\dagger}(\boldsymbol{x})i\boldsymbol{\sigma}\cdot\boldsymbol{D}\Psi_{R}(\boldsymbol{x})\nonumber \\
 & = & \sum_{k_{y},k_{z}}[k_{z}\theta(k_{z})a_{0}^{\dagger}(k_{y},k_{z})a_{0}(k_{y},k_{z})+(-k_{z})\theta(-k_{z})b_{0}^{\dagger}(k_{y},k_{z})b_{0}(k_{y},k_{z})]\nonumber \\
 &  & +\sum_{n,k_{y},k_{z}}E_{n}(k_{z})[a_{n}^{\dagger}(k_{y},k_{z})a_{n}(k_{y},k_{z})+b_{n}^{\dagger}(k_{y},k_{z})b_{n}(k_{y},k_{z})],\label{eq:ss6}
\end{eqnarray}
\begin{eqnarray}
N & = & \int d^{3}x\Psi_{R}^{\dagger}(\boldsymbol{x})\Psi_{R}(\boldsymbol{x})\nonumber \\
 & = & \sum_{k_{y},k_{z}}[\theta(k_{z})a_{0}^{\dagger}(k_{y},k_{z})a_{0}(k_{y},k_{z})-\theta(-k_{z})b_{0}^{\dagger}(k_{y},k_{z})b_{0}(k_{y},k_{z})]\nonumber \\
 &  & \sum_{n,k_{y},k_{z}}[a_{n}^{\dagger}(k_{y},k_{z})a_{n}(k_{y},k_{z})-b_{n}^{\dagger}(k_{y},k_{z})b_{n}(k_{y},k_{z})],\label{eq:ss7}
\end{eqnarray}
where we have dropped the infinite vacuum term. This can be renormalized
in phyiscs calculation and does not affect our result on CME coefficient.
It is clear that $\theta(k_{z})a_{0}^{\dagger}(k_{y},k_{z})a_{0}(k_{y},k_{z})$
and $a_{n}^{\dagger}(k_{y},k_{z})a_{n}(k_{y},k_{z})$ are occupied
number operators of particles for different Landau levels, and $\theta(-k_{z})b_{0}^{\dagger}(k_{y},k_{z})b_{0}(k_{y},k_{z})$
and $b_{n}^{\dagger}(k_{y},k_{z})b_{n}(k_{y},k_{z})$ are occupied
number operators of antiparticles for different Landau levels. Note
that if we had not introduced the two theta functions $\theta(\pm k_{z})$
in front of $a_{0}(k_{y},k_{z})$ and $b_{0}^{\dagger}(k_{y},k_{z})$
in the decomposition of $\Psi_{R}(\boldsymbol{x})$, we would not
do the second quantization procedure successfully.

\section{Chiral magnetic effect}

\label{sec:Chiral-magnetic-effect}Suppose that the system of righthand
ferimons within an external uniform magnetic field $\boldsymbol{B}=B\mathbf{e}_{z}$
is in equilibrium with a reservior with temperature $T$ and chemical
potential $\mu_{R}$. Then the density operator $\hat{\rho}$ for
this righthand fermion system is 
\begin{equation}
\hat{\rho}=\frac{1}{Z}e^{-\beta(H-\mu_{R}N)},\label{eq:tt20}
\end{equation}
where $\beta=1/T$ is the inverse temperature, and $Z$ is the grand
canonical partition function,

\begin{equation}
Z=\text{Tr}\,e^{-\beta(H-\mu_{R}N)}.\label{eq:tt21}
\end{equation}
The expectation value of an operator $\hat{F}$ in the equilibrium
state can be calculated as 
\begin{equation}
\langle:\hat{F}:\rangle=\text{Tr}\,(\hat{\rho}\hat{F}).\label{eq:tt22}
\end{equation}

In Appendix \ref{sec:occupied number} we have calculated the expectation
values of occupied number operators as 
\begin{align}
\langle:\theta(k_{z})a_{0}^{\dagger}(k_{y},k_{z})a_{0}(k_{y},k_{z}):\rangle & =\frac{\theta(k_{z})}{e^{\beta(k_{z}-\mu_{R})}+1}\nonumber \\
\langle:\theta(-k_{z})b_{0}^{\dagger}(k_{y},k_{z})b_{0}(k_{y},k_{z}):\rangle & =\frac{\theta(-k_{z})}{e^{\beta(-k_{z}+\mu_{R})}+1}\nonumber \\
\langle:a_{n}^{\dagger}(k_{y},k_{z})a_{n}(k_{y},k_{z}):\rangle & =\frac{1}{e^{\beta[E_{n}(k_{z})-\mu_{R}]}+1}\nonumber \\
\langle:b_{n}^{\dagger}(k_{y},k_{z})b_{n}(k_{y},k_{z}):\rangle & =\frac{1}{e^{\beta[E_{n}(k_{z})+\mu_{R}]}+1}.\label{eq:tt23}
\end{align}
The macroscopic electric current for righthand fermion system is 
\begin{equation}
\boldsymbol{J}_{R}=\bigg\langle:\Psi_{R}^{\dagger}(\boldsymbol{x})\boldsymbol{\sigma}\Psi_{R}(\boldsymbol{x}):\bigg\rangle.\label{eq:tt24}
\end{equation}

According to the rotational invariance of this system along $z$-axis,
$J_{R}^{x}=J_{R}^{y}=0$. In the following, we will calculate $J_{R}^{z}$.
Taking use of Eq. (\ref{eq:ss2}), we can see 
\begin{align}
J_{R}^{z}= & \langle:\Psi_{R}^{\dagger}(\boldsymbol{x})\sigma^{3}\Psi_{R}(\boldsymbol{x}):\rangle\nonumber \\
= & \sum_{k_{y},k_{z}}\bigg(\langle:\theta(k_{z})a_{0}^{\dagger}(k_{y},k_{z})a_{0}(k_{y},k_{z}):\rangle+\langle:\theta(-k_{z})b_{0}(k_{y},k_{z})b_{0}^{\dagger}(k_{y},k_{z}):\rangle\bigg)\nonumber \\
 & \times\psi_{R0}^{\dagger}(k_{y},k_{z};\boldsymbol{x})\sigma^{3}\psi_{R0}(k_{y},k_{z};\boldsymbol{x})\nonumber \\
 & +\sum_{n,k_{y},k_{z}}\langle:a_{n}^{\dagger}(k_{y},k_{z})a_{n}(k_{y},k_{z}):\rangle\psi_{Rn+}^{\dagger}(k_{y},k_{z};\boldsymbol{x})\sigma^{3}\psi_{Rn+}(k_{y},k_{z};\boldsymbol{x})\nonumber \\
 & +\sum_{n,k_{y},k_{z}}\langle:b_{n}(k_{y},k_{z})b_{n}^{\dagger}(k_{y},k_{z}):\rangle\psi_{Rn-}^{\dagger}(k_{y},k_{z};\boldsymbol{x})\sigma^{3}\psi_{Rn-}(k_{y},k_{z};\boldsymbol{x})\nonumber \\
= & \sum_{k_{y},k_{z}}\bigg(\frac{\theta(k_{z})}{e^{\beta(k_{z}-\mu_{R})}+1}-\frac{\theta(-k_{z})}{e^{\beta(-k_{z}+\mu_{R})}+1}\bigg)\psi_{R0}^{\dagger}(k_{y},k_{z};\boldsymbol{x})\sigma^{3}\psi_{R0}(k_{y},k_{z};\boldsymbol{x})\nonumber \\
 & +\sum_{n,k_{y},k_{z}}\frac{1}{e^{\beta[E_{n}(k_{z})-\mu_{R}]}+1}\psi_{Rn+}^{\dagger}(k_{y},k_{z};\boldsymbol{x})\sigma^{3}\psi_{Rn+}(k_{y},k_{z};\boldsymbol{x})\nonumber \\
 & -\sum_{n,k_{y},k_{z}}\frac{1}{e^{\beta[E_{n}(k_{z})+\mu_{R}]}+1}\psi_{Rn-}^{\dagger}(k_{y},k_{z};\boldsymbol{x})\sigma^{3}\psi_{Rn-}(k_{y},k_{z};\boldsymbol{x}).\label{eq:ss8}
\end{align}
Firstly we sum over $k_{y}$ for $\psi_{R0}^{\dagger}(k_{y},k_{z};\boldsymbol{x})\sigma^{3}\psi_{R0}(k_{y},k_{z};\boldsymbol{x})$
and $\psi_{Rn\lambda}^{\dagger}(k_{y},k_{z};\boldsymbol{x})\sigma^{3}\psi_{Rn\lambda}(k_{y},k_{z};\boldsymbol{x})$
in Eq. (\ref{eq:ss8}), the results are 
\begin{align}
 & \sum_{k_{y}}\psi_{R0}^{\dagger}(k_{y},k_{z};\boldsymbol{x})\sigma^{3}\psi_{R0}(k_{y},k_{z};\boldsymbol{x})\nonumber \\
= & \frac{1}{L^{2}}\sum_{k_{y}}\left(\begin{array}{cc}
\varphi_{0}(\xi), & 0\end{array}\right)\left(\begin{array}{cc}
1 & 0\\
0 & -1
\end{array}\right)\left(\begin{array}{c}
\varphi_{0}(\xi)\\
0
\end{array}\right)\nonumber \\
= & \frac{1}{2\pi L}\int_{-\infty}^{\infty}dk_{y}[\varphi_{0}(\sqrt{eB}x-k_{y}/\sqrt{eB})]^{2}\nonumber \\
= & \frac{eB}{2\pi L},\label{eq:ap22}
\end{align}
and 
\begin{align}
 & \sum_{k_{y}}\psi_{Rn\lambda}^{\dagger}(k_{y},k_{z};\boldsymbol{x})\sigma^{3}\psi_{Rn\lambda}(k_{y},k_{z};\boldsymbol{x})\nonumber \\
= & \frac{1}{2\pi L}\int_{-\infty}^{\infty}dk_{y}c_{n\lambda}^{2}(k_{z})\bigg([\varphi_{n}(\xi)]^{2}-\frac{2neB[\varphi_{n-1}(\xi)]^{2}}{[k_{z}+\lambda E_{n}(k_{z})]^{2}}\bigg)\nonumber \\
= & \frac{eB}{2\pi L}c_{n\lambda}^{2}(k_{z})\bigg(1-\frac{2neB}{[k_{z}+\lambda E_{n}(k_{z})]^{2}}\bigg)\nonumber \\
= & \frac{eB}{2\pi L}c_{n\lambda}^{2}(k_{z})[2-c_{n\lambda}^{-2}(k_{z})]\nonumber \\
= & \frac{eB}{2\pi L}\cdot\frac{\lambda k_{z}}{E_{n}(k_{z})}.\label{eq:ap23}
\end{align}
Secondly, we sum over $k_{z}$ in the third equal sign of Eq. (\ref{eq:ss8}),
\begin{align}
J_{R}^{z}= & \sum_{k_{z}}\bigg(\frac{\theta(k_{z})}{e^{\beta(k_{z}-\mu_{R})}+1}-\frac{\theta(-k_{z})}{e^{\beta(-k_{z}+\mu_{R})}+1}\bigg)\frac{eB}{2\pi L}\nonumber \\
 & +\sum_{n,k_{z}}\bigg(\frac{1}{e^{\beta[E_{n}(k_{z})-\mu_{R}]}+1}+\frac{1}{e^{\beta[E_{n}(k_{z})+\mu_{R}]}+1}\bigg)\frac{eB}{2\pi L}\cdot\frac{k_{z}}{E_{n}(k_{z})}\nonumber \\
= & \frac{eB}{4\pi^{2}}\int_{-\infty}^{\infty}dk_{z}\bigg(\frac{\theta(k_{z})}{e^{\beta(k_{z}-\mu_{R})}+1}-\frac{\theta(-k_{z})}{e^{\beta(-k_{z}+\mu_{R})}+1}\bigg)+0\nonumber \\
= & \frac{eB}{4\pi^{2}}\mu_{R}.\label{eq:ap24}
\end{align}
Combining Eq. (\ref{eq:ap24}) and $J_{R}^{x}=J_{R}^{y}=0$ gives
\begin{equation}
\boldsymbol{J}_{R}=\frac{e\mu_{R}}{4\pi^{2}}\boldsymbol{B}.\label{eq:ap25}
\end{equation}

From the calculation above, we can see that only the lowest Landau
level contributes to Eq. (\ref{eq:ap25}). Similar calculation for
lefthand fermion system shows that 
\begin{equation}
\boldsymbol{J}_{L}=-\frac{e\mu_{L}}{4\pi^{2}}\boldsymbol{B}.\label{eq:ss5}
\end{equation}
Actually, we can also obtain equation (\ref{eq:ss5}) from (\ref{eq:ap25})
under space inversion: $\boldsymbol{J}_{R}\rightarrow-\boldsymbol{J}_{L}$,
$\mu_{R}\rightarrow\mu_{L}$, $\boldsymbol{B}\rightarrow\boldsymbol{B}$.
If the system composes of righthand and lefthand fermions, then the
vector current $\boldsymbol{J}_{V}$ and axial current $\boldsymbol{J}_{A}$
are 
\begin{equation}
\boldsymbol{J}_{V}=\boldsymbol{J}_{R}+\boldsymbol{J}_{L}=\frac{e\mu_{5}}{2\pi^{2}}\boldsymbol{B},\label{eq:tt25}
\end{equation}
\begin{equation}
\boldsymbol{J}_{A}=\boldsymbol{J}_{R}-\boldsymbol{J}_{L}=\frac{e\mu}{2\pi^{2}}\boldsymbol{B},\label{eq:tt26}
\end{equation}
where $\mu_{5}=(\mu_{R}-\mu_{L})/2$ is called chiral chemical potential
and $\mu=(\mu_{R}+\mu_{L})/2$. So far we have derived CME for chiral
fermion system through solving Landau levels. We amphasize that Eqs.
(\ref{eq:tt25}, \ref{eq:tt26}) are valid for any strength of magnetic
field, which is different from the weak magnetic field approximation
through Wigner function approach in Sec. \ref{sec:Wigner-function}.

\section{Physical picture of the lowest Landau level}

\label{sec:Physical-picture}In this section we discuss the physical
picture of the lowest Landau level. The wavefunction and energy of
the lowest Landau level ($n=0$) for righthand fermion field is 
\begin{equation}
\psi_{R0}(k_{y},k_{z};\boldsymbol{x})=\left(\begin{array}{c}
\varphi_{0}\\
0
\end{array}\right)\frac{1}{L}e^{i(yk_{y}+zk_{z})},\ \ E=k_{z}.\label{eq:tt1}
\end{equation}
Setting $k_{y}=0$ in Eq. (\ref{eq:tt1}), we calculate the Hamiltonian,
particle number, $z$-component of momentum, and $z$-component of
spin angular momentum of the righthand fermion system for the lowest
Landau level as follows, 
\begin{align}
H & =\sum_{k_{z}}[k_{z}\theta(k_{z})a_{0}^{\dagger}(0,k_{z})a_{0}(0,k_{z})+(-k_{z})\theta(-k_{z})b_{0}^{\dagger}(0,k_{z})b_{0}(0,k_{z})],\nonumber \\
N & =\sum_{k_{z}}[\theta(k_{z})a_{0}^{\dagger}(0,k_{z})a_{0}(0,k_{z})+(-1)\theta(-k_{z})b_{0}^{\dagger}(0,k_{z})b_{0}(0,k_{z})],\nonumber \\
P_{z} & =\sum_{k_{z}}[k_{z}\theta(k_{z})a_{0}^{\dagger}(0,k_{z})a_{0}(0,k_{z})+(-k_{z})\theta(-k_{z})b_{0}^{\dagger}(0,k_{z})b_{0}(0,k_{z})],\nonumber \\
S_{z} & =\sum_{k_{z}}[\frac{1}{2}\theta(k_{z})a_{0}^{\dagger}(0,k_{z})a_{0}(0,k_{z})+(-\frac{1}{2})\theta(-k_{z})b_{0}^{\dagger}(0,k_{z})b_{0}(0,k_{z})].\label{eq:tt27}
\end{align}
Then we have a picture for the lowest Landau level: The operator $\theta(k_{z})a_{0}^{\dagger}(0,k_{z})$
produces a particle with charge $e$, energy $k_{z}>0$, $z$-component
of momentum $k_{z}>0$, and $z$-component of spin angular momentum
$+\frac{1}{2}$ (helicity $h=+1$); The operator $\theta(-k_{z})b_{0}^{\dagger}(0,k_{z})$
produces a particle with charge $-e$, energy $-k_{z}>0$, $z$-component
of momentum $-k_{z}>0$, and $z$-component of spin angular momentum
$-\frac{1}{2}$ (helicity $h=-1$). This picture means that all righthand
fermions/antifermions move along $(+z)$-axis, with righthand fermions
spinning along $(+z)$-axis and righthand antifermions spinning along
$-z$-axis. If $\mu_{R}>0$, which means righthand fermions are more
than righthand anti-fermions, then there will be net electric current
moving along $(+z)$-axis, which is called chiral magnetic effect
for righthand fermion system.

The analogous analysis can be applied to lefthand fermions. The picture
of the lowest Landau level for a lefthand fermion is: all lefthand
fermions/antifermions move along $(-z)$-axis, with left fermions
spinning along $(+z)$-axis and lefthand antifermions spinning along
$(-z)$-axis. If $\mu_{L}>0$, which means lefthand fermions are more
than lefthand anti-fermions, then there will be net electric current
moving along $(-z)$-axis, which is called chiral magnetic effect
for lefthand fermion system.

Since the total electric current $\boldsymbol{J}_{V}$ of the chiral
fermion system is the summation of the electric current $\boldsymbol{J}_{R}$
of the righthand fermion system and the electric current $\boldsymbol{J}_{L}$
of the lefthand fermion system, whether $\boldsymbol{J}_{V}$ is along
$(+z)$-axis or not will only depend on the sign of $(\mu_{R}-\mu_{L})$.
So we have explained CME for chiral fermion system microscopically.

\section{Summary}

\label{sec:Summary}Chiral magnetic effect (CME) arises from the lowest
Landau level both for massive Dirac fermion system and chiral fermion
system. For massive case, the physical picture of how the lowest Landau
level contributes to CME is not very clear. When we solve the Landau
levels for chiral fermion system in the background of a uniform magnetic
field, by performing the second quantization for the chiral fermion
field, expanding field operator by eigenfunction of Landau levels,
and calculating the ensemble average of vector current operator, we
natrually obtain the equation for CME. We point out that we had not
made any approximation for the strength of magnetic field in the calculation.
It is worth mentioning that we had introduced two theta functions
$\theta(\pm k_{z})$ in front of $a_{0}(k_{y},k_{z})$ and $b_{0}^{\dagger}(k_{y},k_{z})$
in the decomposition of $\Psi_{R}(\boldsymbol{x})$, which is crucial
for performing the subsequent procedure of second quantization successfully.
When we carefully analyze the lowest Landau level, we find that all
righthand (chirality is $+1$) fermions move along positive $z$-direction
and all lefthand (chirality is $-1$) fermions move along negative
$z$-direction, and CME can also be explained microscopically within
this picture of the lowest Landau level.

\section{\textit{Acknowledgments}}

We are grateful to Hai-Cang Ren and Xin-Li Sheng for valuable discussions.
R.-H. F. is supported by the National Natural Science Foundation of
China (NSFC) under Grant No. 11847220. D.-F. H. is in part supported
by the National Natural Science Foundation of China (NSFC) under Grant
Nos. 11735007, 11890711.

\appendix

\section{Vlasov equation and off mass-shell equation}

\label{sec:Vlasov-equation}The quadratic form for the equation of
motion of Wigner function $\mathscr{W}(x,p)$ is 
\begin{equation}
\bigg(K^{2}-\frac{i}{2}\sigma^{\mu\nu}[K_{\mu},K_{\nu}]\bigg)\mathscr{W}=0.\label{eq:oo6-1}
\end{equation}
Taking use of $K^{2}=p^{2}-\frac{1}{4}\nabla^{2}+ip\cdot\nabla$ and
$[K_{\mu},K_{\nu}]=-ieF_{\mu\nu}$, Eq. (\ref{eq:oo6-1}) becomes
\begin{equation}
(p^{2}-\frac{1}{4}\nabla^{2}+ip\cdot\nabla-\frac{1}{2}eF_{\mu\nu}\sigma^{\mu\nu})\mathscr{W}=0.\label{eq:oo7-1}
\end{equation}
Note that $\mathscr{W}$ and $\sigma^{\mu\nu}$ satisfies $\mathscr{W}=\gamma^{0}\mathscr{W}^{\dagger}\gamma^{0}$
and $\sigma^{\mu\nu}=\gamma^{0}\sigma^{\mu\nu\dagger}\gamma^{0}$.
Taking Hermi conjugation and then multiplying $\gamma^{0}$ from both
sides of Eq. (\ref{eq:oo7-1}) yield 
\begin{equation}
(p^{2}-\frac{1}{4}\nabla^{2}-ip\cdot\nabla)\mathscr{W}-\frac{1}{2}eF_{\mu\nu}\mathscr{W}\sigma^{\mu\nu}=0.\label{eq:oo8-1}
\end{equation}
Eq. (\ref{eq:oo7-1}) minus Eq. (\ref{eq:oo8-1}) yields the Vlasov
equation for $\mathscr{W}$, 
\begin{equation}
ip\cdot\nabla\mathscr{W}-\frac{1}{4}eF_{\mu\nu}[\sigma^{\mu\nu},\mathscr{W}]=0.\label{eq:oo10-1}
\end{equation}
Eq. (\ref{eq:oo7-1}) plus Eq. (\ref{eq:oo8-1}) yields the off mass-shell
equation for $\mathscr{W}$, 
\begin{equation}
(p^{2}-\frac{1}{4}\nabla^{2})\mathscr{W}-\frac{1}{4}eF_{\mu\nu}\{\sigma^{\mu\nu},\mathscr{W}\}=0.\label{eq:oo9-1}
\end{equation}
To calculate $[\sigma^{\mu\nu},W]$ and $\{\sigma^{\mu\nu},W\}$ in
Eqs. (\ref{eq:oo10-1}) (\ref{eq:oo9-1}), we list following useful
identities, 
\begin{eqnarray}
[\sigma^{\mu\nu},1] & = & 0,\nonumber \\{}
[\sigma^{\mu\nu},i\gamma^{5}] & = & 0,\nonumber \\{}
[\sigma^{\mu\nu},\gamma^{\rho}] & = & -2ig^{\rho[\mu}\gamma^{\nu]},\nonumber \\{}
[\sigma^{\mu\nu},\gamma^{5}\gamma^{\rho}] & = & -2ig^{\rho[\mu}\gamma^{5}\gamma^{\nu]},\nonumber \\{}
[\sigma^{\mu\nu},\sigma^{\rho\sigma}] & = & 2ig^{\mu[\rho}\sigma^{\sigma]\nu}-2ig^{\nu[\rho}\sigma^{\sigma]\mu},\label{eq:ap29}\\
\nonumber \\
\{\sigma^{\mu\nu},1\} & = & 2\sigma^{\mu\nu},\nonumber \\
\{\sigma^{\mu\nu},i\gamma^{5}\} & = & -\epsilon^{\mu\nu\rho\sigma}\sigma_{\rho\sigma},\nonumber \\
\{\sigma^{\mu\nu},\gamma^{\rho}\} & = & 2\epsilon^{\mu\nu\rho\sigma}\gamma^{5}\gamma_{\sigma},\nonumber \\
\{\sigma^{\mu\nu},\gamma^{5}\gamma^{\rho}\} & = & 2\epsilon^{\mu\nu\rho\sigma}\gamma_{\sigma},\nonumber \\
\{\sigma^{\mu\nu},\sigma^{\rho\sigma}\} & = & 2g^{\mu[\rho}g^{\sigma]\nu}+2i\epsilon^{\mu\nu\rho\sigma}\gamma^{5}.\label{eq:30}
\end{eqnarray}
Then all matrices appearing in Eqs. (\ref{eq:oo10-1}) (\ref{eq:oo9-1})
are the 16 independent $\Gamma$-matrices, whose coefficients must
be zero. These coefficient equations are the Vlasov equations and
the off mass-shell equations for $\mathscr{F},\mathscr{P},\mathscr{V}_{\mu},\mathscr{A}_{\mu},\mathscr{S}_{\mu\nu}$.
The Vlasov equations are 
\begin{eqnarray}
p\cdot\nabla\mathscr{F} & = & 0,\nonumber \\
p\cdot\nabla\mathscr{P} & = & 0,\nonumber \\
p\cdot\nabla\mathscr{V}_{\mu} & = & eF_{\mu\nu}\mathscr{V}^{\nu},\nonumber \\
p\cdot\nabla\mathscr{A}_{\mu} & = & eF_{\mu\nu}\mathscr{A}^{\nu},\nonumber \\
p\cdot\nabla\mathscr{Q}_{\mu\nu} & = & eF_{\ [\mu}^{\rho}\mathscr{Q}_{\nu]\rho},\label{eq:ap30}
\end{eqnarray}
and the off mass-shell equations are 
\begin{eqnarray}
(p^{2}-\frac{1}{4}\nabla^{2})\mathscr{F} & = & \frac{1}{2}eF_{\mu\nu}\mathscr{Q}^{\mu\nu},\nonumber \\
(p^{2}-\frac{1}{4}\nabla^{2})\mathscr{P} & = & \frac{1}{2}e\tilde{F}_{\mu\nu}\mathscr{Q}^{\mu\nu},\nonumber \\
(p^{2}-\frac{1}{4}\nabla^{2})\mathscr{V}_{\mu} & = & -e\tilde{F}_{\mu\nu}\mathscr{A}^{\nu},\nonumber \\
(p^{2}-\frac{1}{4}\nabla^{2})\mathscr{A}_{\mu} & = & -e\tilde{F}_{\mu\nu}\mathscr{V}^{\nu},\nonumber \\
(p^{2}-\frac{1}{4}\nabla^{2})\mathscr{Q}_{\mu\nu} & = & e(F_{\mu\nu}\mathscr{F}-\tilde{F}_{\mu\nu}\mathscr{P}),\label{eq:ap31}
\end{eqnarray}
where $\tilde{F}_{\mu\nu}=\frac{1}{2}\varepsilon_{\mu\nu\rho\sigma}F^{\rho\sigma}$.

\section{Landau levels for righthand feimion field}

\label{sec:Landau levels-1}Now we will solve following eigenvalue
equation in detail, 
\begin{equation}
i\boldsymbol{\sigma}\cdot\boldsymbol{D}\psi_{R}(\boldsymbol{x})=E\psi_{R}(\boldsymbol{x}),\label{eq:app1}
\end{equation}
with $\boldsymbol{D}=(-\partial_{x},-\partial_{y}+ieBx,-\partial_{z})$.
Since the operator $i\boldsymbol{\sigma}\cdot\boldsymbol{D}$ is commutative
with $\hat{p}_{y}=-i\partial_{y},\hat{p}_{z}=-i\partial_{z}$, then
we can choose $\psi_{R}$ as the commom eigenstate of $i\boldsymbol{\sigma}\cdot\boldsymbol{D}$,
$\hat{p}_{y}$ and $\hat{p}_{z}$ as follows 
\begin{equation}
\psi_{R}(x,y,z)=\left(\begin{array}{c}
\phi_{1}(x)\\
\phi_{2}(x)
\end{array}\right)\frac{1}{L}e^{i(yk_{y}+zk_{z})},\label{eq:ap2}
\end{equation}
where $L$ is the length of the system in $y$- and $z$- directions.
The explicit form of $\boldsymbol{\sigma}\cdot\boldsymbol{D}$ is
\begin{equation}
\boldsymbol{\sigma}\cdot\boldsymbol{D}=\left(\begin{array}{cc}
-\partial_{z} & -\partial_{x}+i\partial_{y}+eBx\\
-\partial_{x}-i\partial_{y}-eBx & \partial_{z}
\end{array}\right).\label{eq:ap3}
\end{equation}

Putting Eq. (\ref{eq:ap2}) (\ref{eq:ap3}) into Eq. (\ref{eq:app1}),
we obtain the group of differential equations for $\phi_{1}(x)$ and
$\phi_{2}(x)$ as 
\begin{eqnarray}
i(k_{z}-E)\phi_{1}+(\partial_{x}+k_{y}-eBx)\phi_{2} & = & 0,\label{eq:1a}\\
(\partial_{x}-k_{y}+eBx)\phi_{1}-i(k_{z}+E)\phi_{2} & = & 0.\label{eq:1b}
\end{eqnarray}
From Eq. (\ref{eq:1b}) we can express $\phi_{2}$ by $\phi_{1}$,
then Eq. (\ref{eq:1a}) becomes 
\begin{equation}
\partial_{x}^{2}\phi_{1}+\bigg(E^{2}+eB-k_{z}^{2}-e^{2}B^{2}(x-\frac{k_{y}}{eB})^{2}\bigg)\phi_{1}=0,\label{eq:1d}
\end{equation}
which is a typical harmonic oscillator equation. Define a dimensionless
variable $\xi=\sqrt{eB}(x-k_{y}/eB)$, and $\phi_{1}(x)=\varphi(\xi)$,
then (\ref{eq:1d}) becomes 
\begin{equation}
\frac{d^{2}\varphi}{d\xi^{2}}+\bigg(\frac{E^{2}-k_{z}^{2}}{eB}+1-\xi^{2}\bigg)\varphi=0.\label{eq:ss1}
\end{equation}
With the boundary condition $\varphi\rightarrow0$ as $\xi\rightarrow\pm\infty$,
we must set 
\begin{equation}
\frac{E^{2}-k_{z}^{2}}{eB}+1=2n+1,\label{eq:ap4}
\end{equation}
with $n=0,1,2,\cdots$. So energy $E$ can only take following discrete
values, 
\begin{equation}
E=\pm E_{n}(k_{z})\equiv\pm\sqrt{2neB+k_{z}^{2}},\label{eq:ap5}
\end{equation}
where we have defined $E_{n}(k_{z})=\sqrt{2neB+k_{z}^{2}}$. The corresponding
normalised solution for equation (\ref{eq:1d}) is 
\begin{equation}
\phi_{1}(x)=\varphi_{n}(\xi)=N_{n}e^{-\xi^{2}/2}H_{n}(\xi),\label{eq:ap6}
\end{equation}
where $N_{n}=(eB)^{\frac{1}{4}}\pi^{-\frac{1}{4}}(2^{n}n!)^{-\frac{1}{2}}$,
and $H_{n}(\xi)=(-1)^{n}e^{\xi^{2}}\frac{d^{n}}{d\xi^{n}}e^{-\xi^{2}}$.
For energy $E=\lambda E_{n}(k_{z})$ ($\lambda=\pm1$), we can obtain
$\phi_{2}$ as 
\begin{equation}
\phi_{2}(x)=\frac{\sqrt{eB}(\partial_{\xi}+\xi)\varphi_{n}(\xi)}{i(k_{z}+E)}=\frac{i[k_{z}-\lambda E_{n}(k_{z})]}{\sqrt{2neB}}\varphi_{n-1}(\xi),\label{eq:ap7}
\end{equation}
where we have used $(\partial_{\xi}+\xi)\varphi_{n}(\xi)=\sqrt{2n}\varphi_{n-1}(\xi)$.
Define $F_{n\lambda}(k_{z})=[k_{z}-\lambda E_{n}(k_{z})]/\sqrt{2neB}$,
then the eigenfunction with eigenvalue $E=\lambda E_{n}(k_{z})$ is

\begin{equation}
\psi_{Rn\lambda}(k_{y},k_{z};\boldsymbol{x})=\left(\begin{array}{c}
\varphi_{n}(\xi)\\
iF_{n\lambda}(k_{z})\varphi_{n-1}(\xi)
\end{array}\right)\frac{1}{L}e^{i(yk_{y}+zk_{z})}.\label{eq:ap8}
\end{equation}
It is very subtle when $n=0$ in Eq. (\ref{eq:ap7}). When $n=0,E=k_{z}$,
the first equal sign of Eq. (\ref{eq:ap7}) indicates $\phi_{2}=0$
due to $(\partial_{\xi}+\xi)\varphi_{0}(\xi)=0$. Then the corresponding
eigenfunction becomes 
\begin{equation}
\psi_{R0}(k_{y},k_{z};\boldsymbol{x})=\left(\begin{array}{c}
\varphi_{0}(\xi)\\
0
\end{array}\right)\frac{1}{L}e^{i(yk_{y}+zk_{z})}.\label{eq:ap9}
\end{equation}
When $n=0,E=-k_{z}$, the denominator of the first equal sign of Eq.
(\ref{eq:ap7}) becomes zero, in which case we must directly deal
with Eqs. (\ref{eq:1a}) (\ref{eq:1b}). In this case Eqs. (\ref{eq:1a})
(\ref{eq:1b}) become 
\begin{eqnarray}
2ik_{z}\phi_{1}+(\partial_{x}+k_{y}-eBx)\phi_{2} & = & 0,\label{eq:1h-1}\\
(\partial_{x}-k_{y}+eBx)\phi_{1} & = & 0.\label{eq:1g-1}
\end{eqnarray}
Eq. (\ref{eq:1g-1}) gives $\phi_{1}(x)\sim\exp[-\frac{1}{2}eBx^{2}+xk_{y}]$,
then Eq. (\ref{eq:1h-1}) becomes 
\begin{equation}
2ik_{z}\exp\bigg(-\frac{1}{2}eBx^{2}+xk_{y}\bigg)+(\partial_{x}+k_{y}-eBx)\phi_{2}=0.\label{eq:1i-1}
\end{equation}
When $x\rightarrow\pm\infty$, Eq. (\ref{eq:1i-1}) tends to 
\begin{equation}
(\partial_{x}-eBx)\phi_{2}=0,\label{eq:ap10}
\end{equation}
whose solution is $\phi_{2}\sim\exp(\frac{1}{2}eBx^{2})$ which is
divergent as $x\rightarrow\pm\infty$. So there exits no physical
solution when $n=0,E=-k_{z}$.

So far we obtain the eigenfunctions and eigenvalues of the Hamiltonian
of the righthand fermion field as follows:

For $n=0$ Landau level, the wavefunction with energy $E=k_{z}$ is

\begin{equation}
\psi_{R0}(k_{y},k_{z};\boldsymbol{x})=\left(\begin{array}{c}
\varphi_{0}\\
0
\end{array}\right)\frac{1}{L}e^{i(yk_{y}+zk_{z})}.\label{eq:ap11}
\end{equation}

For $n>0$ Landau level, the wavefunction with energy $E=\lambda E_{n}(k_{z})$
are

\begin{equation}
\psi_{Rn\lambda}(k_{y},k_{z};\boldsymbol{x})=c_{n\lambda}\left(\begin{array}{c}
\varphi_{n}\\
iF_{n\lambda}\varphi_{n-1}
\end{array}\right)\frac{1}{L}e^{i(yk_{y}+zk_{z})},\label{eq:ap12}
\end{equation}
where $\lambda=\pm1$, $E_{n}(k_{z})=\sqrt{2neB+k_{z}^{2}}$, $F_{n\lambda}(k_{z})=[k_{z}-\lambda E_{n}(k_{z})]/\sqrt{2neB}$,
$|c_{n\lambda}|^{2}=1/(1+F_{n\lambda}^{2})$.

\section{Expectation value of occupied number operators }

\label{sec:occupied number}In the following, we will calculate the
expectation values of particle number operators. From the expression
of Hamiltonian and total particle number operator in Eqs. (\ref{eq:ss6})
(\ref{eq:ss7}), we can easily get following commutative relations,
\begin{eqnarray}
[N,\theta(k_{z})a_{0}^{\dagger}(k_{y},k_{z})] & = & \theta(k_{z})a_{0}^{\dagger}(k_{y},k_{z})\nonumber \\{}
[N,\theta(-k_{z})b_{0}^{\dagger}(k_{y},k_{z})] & = & -\theta(-k_{z})b_{0}^{\dagger}(k_{y},k_{z})\nonumber \\{}
[N,a_{n}^{\dagger}(k_{y},k_{z})] & = & a_{n}^{\dagger}(k_{y},k_{z})\nonumber \\{}
[N,b_{n}^{\dagger}(k_{y},k_{z})] & = & -b_{n}^{\dagger}(k_{y},k_{z}),\label{eq:ap13}
\end{eqnarray}
\begin{eqnarray}
[H,\theta(k_{z})a_{0}^{\dagger}(k_{y},k_{z})] & = & k_{z}\theta(k_{z})a_{0}^{\dagger}(k_{y},k_{z})\nonumber \\{}
[H,\theta(-k_{z})b_{0}^{\dagger}(k_{y},k_{z})] & = & (-k_{z})\theta(-k_{z})b_{0}^{\dagger}(k_{y},k_{z})\nonumber \\{}
[H,a_{n}^{\dagger}(k_{y},k_{z})] & = & E_{n}(k_{z})a_{n}^{\dagger}(k_{y},k_{z})\nonumber \\{}
[H,b_{n}^{\dagger}(k_{y},k_{z})] & = & E_{n}(k_{z})b_{n}^{\dagger}(k_{y},k_{z}),\label{eq:ap14}
\end{eqnarray}
where we have used $[AB,C]=A\{B,C\}-\{A,C\}B$. Define 
\begin{eqnarray}
\theta(k_{z})a_{0}^{\dagger}(k_{y},k_{z};\beta) & = & e^{-\beta(H-\mu_{R}N)}\theta(k_{z})a_{0}^{\dagger}(k_{y},k_{z})e^{\beta(H-\mu_{R}N)}\nonumber \\
\theta(-k_{z})b_{0}^{\dagger}(k_{y},k_{z};\beta) & = & e^{-\beta(H-\mu_{R}N)}\theta(-k_{z})b_{0}^{\dagger}(k_{y},k_{z})e^{\beta(H-\mu_{R}N)}\nonumber \\
a_{n}^{\dagger}(k_{y},k_{z};\beta) & = & e^{-\beta(H-\mu_{R}N)}a_{n}^{\dagger}(k_{y},k_{z})e^{\beta(H-\mu_{R}N)}\nonumber \\
b_{n}^{\dagger}(k_{y},k_{z};\beta) & = & e^{-\beta(H-\mu_{R}N)}b_{n}^{\dagger}(k_{y},k_{z})e^{\beta(H-\mu_{R}N)}.\label{eq:ap15}
\end{eqnarray}
For $\theta(k_{z})a_{0}^{\dagger}(k_{y},k_{z};\beta)$, we can see

\begin{eqnarray}
\frac{\partial}{\partial\beta}[\theta(k_{z})a_{0}^{\dagger}(k_{y},k_{z};\beta)] & = & -[H-\mu_{R}N,\theta(k_{z})a_{0}^{\dagger}(k_{y},k_{z};\beta)]\nonumber \\
 & = & -e^{-\beta(H-\mu_{R}N)}[H-\mu_{R}N,\theta(k_{z})a_{0}^{\dagger}(k_{y},k_{z})]e^{\beta(H-\mu_{R}N)}\nonumber \\
 & = & -e^{-\beta(H-\mu_{R}N)}[(k_{z}-\mu_{R})\theta(k_{z})a_{0}^{\dagger}(k_{y},k_{z})]e^{\beta(H-\mu_{R}N)}\nonumber \\
 & = & -(k_{z}-\mu_{R})[\theta(k_{z})a_{0}^{\dagger}(k_{y},k_{z};\beta)],\label{eq:ap16}
\end{eqnarray}
with the boundary condition $\theta(k_{z})a_{0}^{\dagger}(k_{y},k_{z};0)=\theta(k_{z})a_{0}^{\dagger}(k_{y},k_{z})$,
which implies 
\begin{equation}
\theta(k_{z})a_{0}^{\dagger}(k_{y},k_{z};\beta)=\theta(k_{z})a_{0}^{\dagger}(k_{y},k_{z})e^{-\beta(k_{z}-\mu_{R})}.\label{eq:ap17}
\end{equation}
Similarly we can obtain 
\begin{eqnarray}
\theta(-k_{z})b_{0}^{\dagger}(k_{y},k_{z};\beta) & = & \theta(-k_{z})b_{0}^{\dagger}(k_{y},k_{z})e^{-\beta(-k_{z}+\mu_{R})}\nonumber \\
a_{n}^{\dagger}(k_{y},k_{z};\beta) & = & a_{n}^{\dagger}(k_{y},k_{z})e^{-\beta[E_{n}(k_{z})-\mu_{R}]}\nonumber \\
b_{n}^{\dagger}(k_{y},k_{z};\beta) & = & b_{n}^{\dagger}(k_{y},k_{z})e^{-\beta[E_{n}(k_{z})+\mu_{R}]}\label{eq:ap18}
\end{eqnarray}

Now we calculate the expectation value of $\langle:\theta(k_{z})a_{0}^{\dagger}(k_{y},k_{z})a_{0}(k_{y},k_{z}):\rangle$.
We can see 
\begin{align}
 & \langle:\theta(k_{z})a_{0}^{\dagger}(k_{y},k_{z})a_{0}(k_{y},k_{z}):\rangle\nonumber \\
= & \text{Tr}\,[\rho\theta(k_{z})a_{0}^{\dagger}(k_{y},k_{z})a_{0}(k_{y},k_{z})]\nonumber \\
= & \frac{1}{Z}\text{Tr}\,\bigg(\theta(k_{z})a_{0}^{\dagger}(k_{y},k_{z};\beta)e^{-\beta(H-\mu_{R}N)}a_{0}(k_{y},k_{z})\bigg)\nonumber \\
= & \frac{1}{Z}\text{Tr}\,\bigg(\theta(k_{z})a_{0}(k_{y},k_{z})a_{0}^{\dagger}(k_{y},k_{z};\beta)e^{-\beta(H-\mu_{R}N)}\bigg)\nonumber \\
= & \langle:\theta(k_{z})a_{0}(k_{y},k_{z})a_{0}^{\dagger}(k_{y},k_{z};\beta):\rangle\nonumber \\
= & \langle:\theta(k_{z})a_{0}(k_{y},k_{z})a_{0}^{\dagger}(k_{y},k_{z}):\rangle e^{-\beta(k_{z}-\mu_{R})}\nonumber \\
= & \theta(k_{z})e^{-\beta(k_{z}-\mu_{R})}-\langle:\theta(k_{z})a_{0}^{\dagger}(k_{y},k_{z})a_{0}(k_{y},k_{z}):\rangle e^{-\beta(k_{z}-\mu_{R})},\label{eq:ap19}
\end{align}
so we obtain

\begin{equation}
\langle\theta(k_{z})a_{0}^{\dagger}(k_{y},k_{z})a_{0}(k_{y},k_{z})\rangle=\frac{\theta(k_{z})}{e^{\beta(k_{z}-\mu_{R})}+1}.\label{eq:ap20}
\end{equation}
Similar calculations obtain 
\begin{align}
\langle\theta(-k_{z})b_{0}^{\dagger}(k_{y},k_{z})b_{0}(k_{y},k_{z})\rangle & =\frac{\theta(-k_{z})}{e^{\beta(-k_{z}+\mu_{R})}+1}\nonumber \\
\langle a_{n}^{\dagger}(k_{y},k_{z})a_{n}(k_{y},k_{z})\rangle & =\frac{1}{e^{\beta[E_{n}(k_{z})-\mu_{R}]}+1}\nonumber \\
\langle b_{n}^{\dagger}(k_{y},k_{z})b_{n}(k_{y},k_{z})\rangle & =\frac{1}{e^{\beta[E_{n}(k_{z})+\mu_{R}]}+1}.\label{eq:ap21}
\end{align}

 \bibliographystyle{apsrev}
\addcontentsline{toc}{section}{\refname}\bibliography{ref-1}

\end{document}